\documentclass[namedreferences]{solarphysics}
\usepackage[optionalrh]{spr-sola-addons} 
\usepackage{epsfig}          
\usepackage{graphicx}        
\usepackage{courier}         
\usepackage{amssymb}        
\usepackage{color}           
\usepackage{url}             
\usepackage[pdfborder={0 0 0 },urlcolor=blue,breaklinks]{hyperref}
\ifx \arxivurl  \undefined \def \arxivurl#1{\href{http://arxiv.org/abs/#1}{\textsf{#1}}}\fi 
\ifx \doiurl    \undefined \def \doiurl#1{\href{http://dx.doi.org/#1}{\textsf{#1}}}\fi 
\ifx \adsurl    \undefined \def \adsurl#1{\href{http://adsabs.harvard.edu/abs/#1}{\textsf{#1}}}\fi 

\newcommand{\adv}{    {\it Adv. Space Res.}}

\newcommand{\aap}{    {\it Astron. Astrophys.}}

\newcommand{\apj}{    {\it Astrophys. J.}}
\newcommand{\apjl}{   {\it Astrophys. J. Lett.}}

\newcommand{\mnras}{  {\it Mon. Not. Roy. Astron. Soc.}}

\newcommand{\solphys}{{\it Solar Phys.}}

\begin{document}

\begin{article}

\begin{opening}

\title{Migration and Extension of Solar Active Longitudinal Zones}

\author{N.~\surname{Gyenge}\sep
        T.~\surname{Baranyi}\sep
        A.~\surname{Ludm\'any}
       }

\runningauthor{N. Gyenge {\it et al.}}
\runningtitle{Active Longitudes}

   \institute{Heliophysical Observatory, Research Centre for Astronomy and Earth Sciences, Hungarian Academy of Sciences, 
\\4010 Debrecen, P.O. Box 30, Hungary \\email: \url{gyenge.norbert@csfk.mta.hu} 
email: \url{baranyi.tunde@csfk.mta.hu} email: \url{ludmany.andras@csfk.mta.hu}}

\begin{abstract}

Solar active longitudes show a characteristic migration pattern in the Carrington coordinate system when they can be identified at all. By following this migration, the longitudinal activity distribution around the center of the band can be determined. The halfwidth of the distribution is found to be varying in Cycles 21\,--\,23, and in some time intervals it was as narrow as 20\,--\,30 degrees. It was more extended around maximum but it was also narrow when the activity jumped to the opposite longitude. Flux emergence exhibited a quasi-periodic variation within the active zone with a period of about 1.3 years. The path of the active longitude migration does not support the view that it might be associated with the 11-year solar cycle. These results were obtained for a limited time interval of a few solar cycles and, bearing in mind uncertainties of the migration path definition, are only indicative. For the major fraction of dataset no systematic active longitudes were found. Sporadic migration of active longitudes was identified only for Cycles 21\,--\,22 in the northern hemisphere and Cycle 23 in the southern hemisphere.

\end{abstract}
\end{opening}

\section{Introduction}
           \label{S-Introduction}

The spatial distribution of active region emergence has been investigated since the creation of the Carrington coordinate system. The equatorward latitudinal migration of sunspot group emergence was first observed by Carrington in 1859, a phenomenon later named after Sp\"orer, and Carrington was also the first to observe active region emergences at the same longitudes (\opencite{Carrington1863}). Since then numerous works have been devoted to the longitudinal grouping of activity. Different terminology has been used but the aim was always to reveal whether the emergence is equally probable at any longitude, or if not, what is the extent of deviation from the axial symmetry is and where are the locations of higher-than-average activity. This is a much more difficult challenge than the study of latitudinal patterns because one cannot assume that the location of enhanced activity is bound to the Carrington system. The diversity of results obtained can partly be explained by the differences of the methods applied, pre-assumptions, input data and time intervals. 

Several attempts have been made to identify the rotation rate of the frame to which the sources of enhanced activity can be bound. As an example, in the most simple approach \inlinecite{Bogart82} makes an autocorrelation analysis on a 128\,--\,year long dataset of Wolf-numbers. He reports two peaks at 27.5 days and 13.6 days, which allows the interpretation that the non-axisymmetric frame has two opposite maxima and rotates with a period of 27.5 days. The procedure results in different values for the individual cycles. This is a Sun-as-a-star method, which has also been used by \inlinecite{Balthasar07}, who made an FFT analysis on the Wolf-numbers and also found peaks around 27.36 and 27.49 days over the 1848\,--\,2006 period, and also at 12.07 days (shorter than half of the longer periods), but the period varies from cycle to cycle and also depends on the cycle phase. 

This approach using a single daily parameter is of course fairly restricted, and the authors admit that a better understanding requires spatial resolution. Nevertheless, this is the method closest to the possibilities of stellar magnetic activity research, {\it i.e.} to see what can be learned from spatially unresolved data. At the same time \inlinecite{Henney05} point out that suggestions of long-term persistency of active longitudes obtained by the method of \inlinecite{Bogart82} may arise by chance. 

The methods considering spatial information are mostly based on sunspot catalogues, primarily on the Greenwich Photoheliographic Results (GPR: 1874\,--\,1976) but also on magnetograms and flare positions. The diversity of the results from different teams is partially caused by the differences in spatial resolution of the data and methods used. Another type of difference is related to the length and date of the chosen time interval. By considering short time intervals one should assume that the rotation rate of the frame containing the active longitudes and the measure of its non-axisymmetry remain constant in time, however, this is not necessarily the case. We will return to the specific results in the discussions by comparing them with the recent findings.  

A possible source of controversy is the use of different pre-assumptions about the geometry and dynamics of the frame carrying the active longitudes. The most sophisticated conception is put forward by \inlinecite{Berdyugina03}, \inlinecite{Usoskin05} and \inlinecite{Berdyugina06}. They assumed a similar differential rotation profile as that observed on the surface and searched for its most probable constants, the equatorial angular velocity and the shear of the profile, {\it i.e.} the constant of the $\sin^2 $ term. They reported a century-scale persistence of active longitudes migrating forward with respect to the Carrington frame and containing two preferred activity regions on the opposite sides of the Sun along with a cyclic modulation that originates from differential rotation, {\it i.e.} from a dependence on the varying latitude. The relative activity levels of the two preferred domains were varying with a mean period of 3.7 years. This variation was interpreted as a solar counterpart of the flip-flop events observed in FK Com \cite{Jetsu93,Olah06}.

The method of these works has been criticized by \inlinecite{Pelt05} and \inlinecite{Pelt06}. They pointed out that certain elements of the above procedures and the considered parameter space may result in a bimodal distribution, {\it i.e.} two persistent preferred longitudes, flip-flop like events and forward migration even on a computer-generated random dataset. These analyses are a warning for our perspective, we should approach the issue with no pre-assumptions about any internal structures. For instance, the assumption of differential rotation implies a cyclic dependence that cannot be regarded as an {\it a-priori} fact. On the other hand, according to \inlinecite{Bigazzi04} the differential rotation makes the non-axisymmetry disappear; it can only remain at the bottom of the convective zone. An apparent signature of the differential rotation can also be detected even in the case of the rigid rotation of a non-axisymmetric magnetic field caused by a stroboscopic effect as demonstrated by \inlinecite{Berdyugina06}.

A possible internal non-axisymmetric magnetic structure would have important theoretical consequences. Following Cowling's (1945) idea about a possible relic magnetic field within the radiative zone, \inlinecite{Kitchatinov05} and \inlinecite{Olemskoy07} searched for a signature of a period of 28.8 days, which is the rotation period of the radiative zone \cite{Schou98}. Their significant value was slightly shorter (28.15 days), but it was only detectable in odd cycles. The authors argue that this seems to support the existence of an off-axis relic field, which could alternately be parallel or antiparallel to the varying field in the convective zone. These authors used GPR-data and considered a sector structure with varying rotation rate. \inlinecite{Plyusnina10} used a similar procedure and obtained different values for the growth and decline phases of the cycles (27.965 and 28.694 days respectively). In these cases, the longitudinal distribution was unimodal. These procedures used pre-defined sector structures without differential rotation.

There are further results for the synodic rotation period of a rigid non-axisymmetric frame. From flare positions 26.72 and 26.61 days \cite{Bai88}, 27.41 days in Cycles 19\,--\,21 \cite{Bai03}, and 22.169 days \cite{Jetsu97} have been found. The radiation zone rotation period of 28.8 days has been found on synoptic maps of photospheric magnetic fields with special filtering techniques by \inlinecite{Mordvinov04}. However, the different rotation periods do not necessarily suggest mistakes or methodological artifacts; some of the differences may be explained by a varying migration pattern as demonstrated by \inlinecite{Juckett06}.

\section{Data and Analysis}
   \label{data}

	\subsection{Activity Maps and Migration Paths} 
  	    \label{paths}
	
The source of observational data that we use is the Debrecen Photoheliographic Data sunspot catalogue (DPD: \opencite{Gyori11}). This material is the continuation of the classic Greenwich Photoheliographic Results (GPR), the source of numerous works in this field. At present, the DPD covers the entire post-Greenwich era: 1977\,--\,2012.  The present study covers the time interval 1979\,--\,2010, 32 years, 416 Carrington Rotations. There are investigations based on longer datasets, but this is the most detailed sunspot catalogue and our present aim is to investigate the dynamics of the identified active longitude zones.

The task is to determine the amount of activity concentration in certain longitudinal belts. The total sunspot area [$ A_{i}$] of all sunspot groups has been computed within $10^{\circ}$ longitudinal bins of the Carrington system for each Carrington Rotation between the years 1977 and 2011 in such a way that the total areas and positions of the individual sunspot groups were taken into account at the time of their largest observed area. Then the data of each bin were divided by the total sum of sunspot areas observed in that rotation in all bins. This normalized quantity represents the activity weight in the given rotation in each bin. The weight of the $i^\mathrm{th}$ bin is:

\begin{eqnarray}
      W_{i} = \frac{A_{i}}{ \sum_{j=1}^{36} A_{j} }
\end{eqnarray}

The representation by normalized weights was criticised by \inlinecite{Pelt06} who argued that this amplifies the role of sunspot groups during low-activity intervals. However, in our case the aim is to find a measure for the activity concentration in certain longitudinal regions independently from the particular level of current solar activity. As far as the position is concerned, the location of emergence was used by \inlinecite{Usoskin05} and \inlinecite{Zhang11}.

\begin{figure}
	\centerline{\includegraphics[scale=1.35, angle=90]{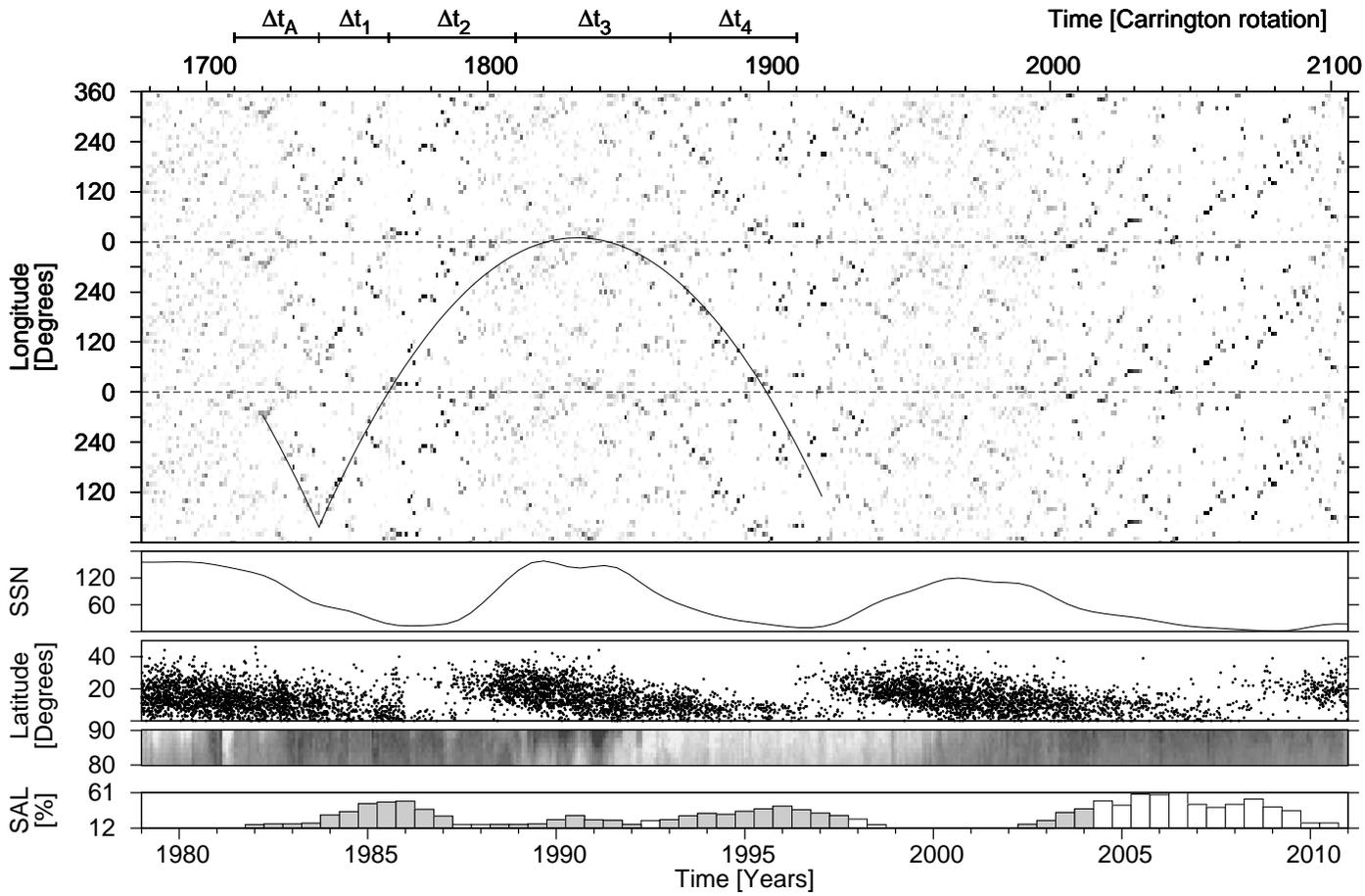}}

              \caption{Longitudinal distribution of solar activity between 1997 and 2011 in the northern hemisphere (first row), the time profiles of Cycles 21\,--\,24 (second row), the hemispheric Sp\"orer-diagram (third row), the polarity of the magnetic dipole field at the northern pole (fourth row). The columns in the last row show the Strength of Active Longitude (SAL), the semiannual means of activity concentration expressed in percentage; grey and empty columns mark smooth and chaotic displacements of the active longitudinal zones respectively, see Section~\ref{path}.}

   \label{north}
   \end{figure}

\begin{figure}
   \centerline{\includegraphics[scale=1.35, angle=90]{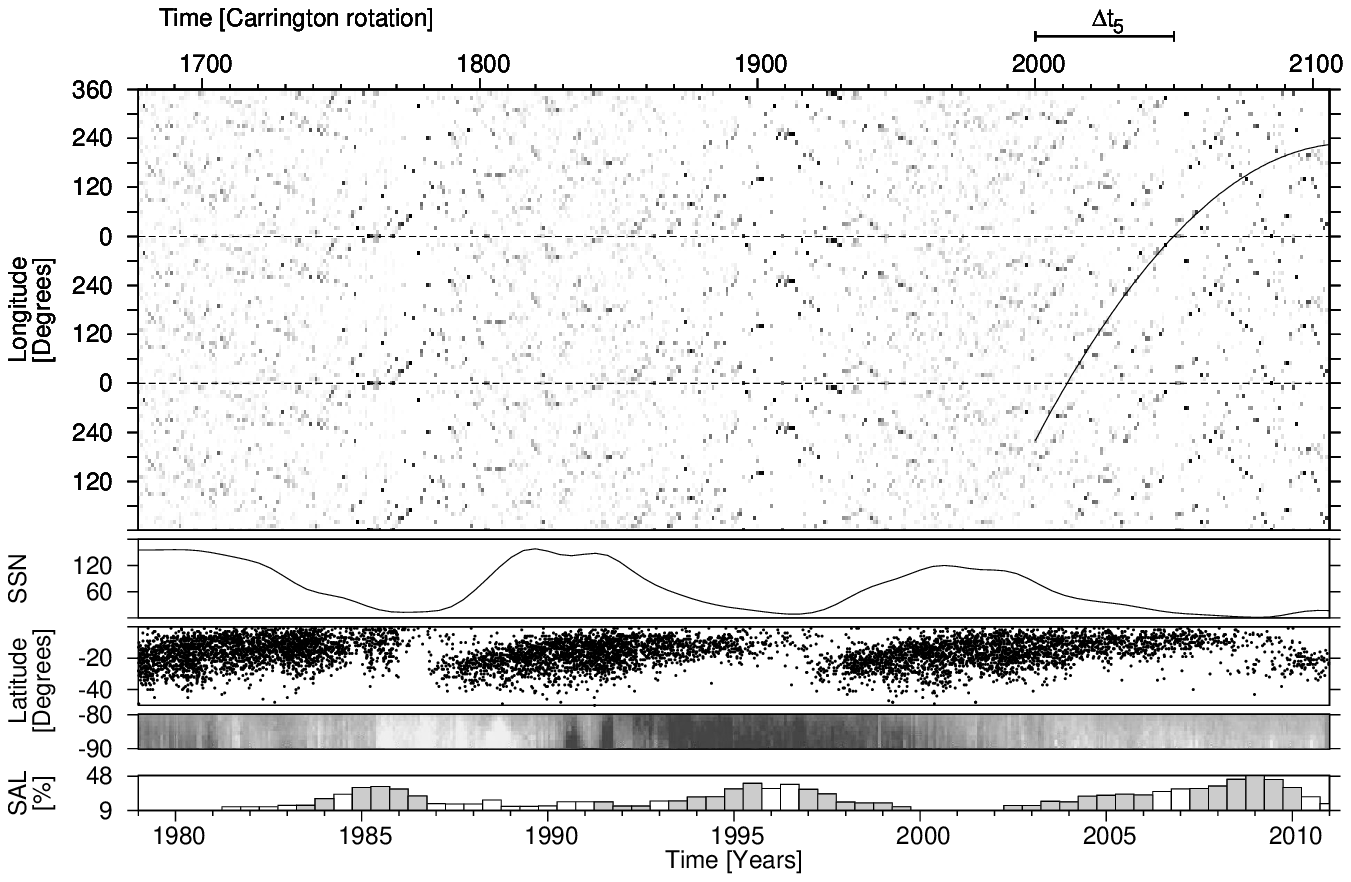}}

              \caption{Longitudinal distribution of solar activity between 1997 and 2011 in the southern hemisphere (first row), the time profiles of Cycles 21\,--\,24 (second row), the hemispheric Sp\"orer-diagram (third row), the polarity of the magnetic dipole field at the southern pole (fourth row). The columns in the last row show the Strength of Active Longitude (SAL), the semiannual means of activity concentration expressed in percentage; grey and empty columns mark smooth and chaotic displacements of the active longitudinal zones respectively, see Section~\ref{path}.}

   \label{south}
   \end{figure}

Figures~\ref{north} and ~\ref{south} show the $W_{i}$ values plotted in each Carrington Rotation ($x$-axis) and $10^{\circ}$ longitudinal bin ($y$-axis) between 1977\,--\,2011 coded with the darkness of grey color for the northern and southern hemispheres. In the $y$-axis the $360^{\circ}$ solar circumference has been repeated three times, similarly to Juckett (2006), in order to follow the occasional shifts of the domains of enhanced sunspot activity in the Carrington system. Below the diagrams the time profiles of Cycles 21\,--\,24 are plotted by using the smoothed International Sunspot Number \cite{SIDC12} as well as the hemispheric Sp\"orer-diagrams by using the sunspot group data of DPD. The next lines of the two figures show the negative and positive magnetic polarities at the poles indicated by dark and light stripes respectively \cite{Hathaway10}. The diagrams of the lowest lines in both figures will be explained in Section~\ref{path}.

Remarkable features of Figures~\ref{north} and ~\ref{south} are the migration tracks of enhanced activity toward both increasing and decreasing longitudes. In Cycles 21\,--\,22 the northern hemisphere exhibits a more expressed path; although around the maximum of Cycle 22 the pixels are fainter because the location of activity is more extended, nevertheless the arc of the return can be recognized. In the southern hemisphere the path is more ambiguous in Cycles 21\,--\,22, it is more evident in Cycles 23\,--\,24, although no receding part can be seen as yet. The continuous curves indicate parabolae that have been fitted on the paths in the following way. The time intervals of most conspicuous migration tracks were selected, in the northern hemisphere: January 1984 \,--\, October 1986 (Carrington Rotation: 1744\,--\,1781) for the advancing migration, November 1992 \,--\, December 1996  (Carrington Rotation: 1862\,--\,1918) for the receding migration; for the southern hemisphere: April 2003 \,--\, January 2007 (Carrington Rotation: 2002\,--\,2054). The longitudinal center of mass of the sunspot-area data has been determined for each rotation in these intervals, {\it i.e.} the mean longitude weighted by the sunspot area. To this set of points a parabola has been fitted by using the standard least squares procedure. The equations of the parabolae are as follows. For the northern hemisphere (Figure~\ref{north}):  $ l=-K(r-1834)^{2}+730  $  where the Carrington longitude is $ L= l \bmod 360^{\circ} $ and {\it r} is the number of Carrington Rotations; 1834 is the Carrington Rotation number at the position of the symmetry axis of the parabola (September 1990). {\it K} is a scale factor between the axis of longitude and the axis of time, $ K$= 0.082 degrees/rotations. For the southern hemisphere (Figure~\ref{south}) $l=-K(r-2120)^{2}+948$, and  $ K$=0.057 degrees/rotation.

\subsection{Widths of Active Belts} 
  	    \label{widths}

\begin{figure}
 \centerline{\hspace*{0.015\textwidth}
               \includegraphics[width=0.515\textwidth,clip=]{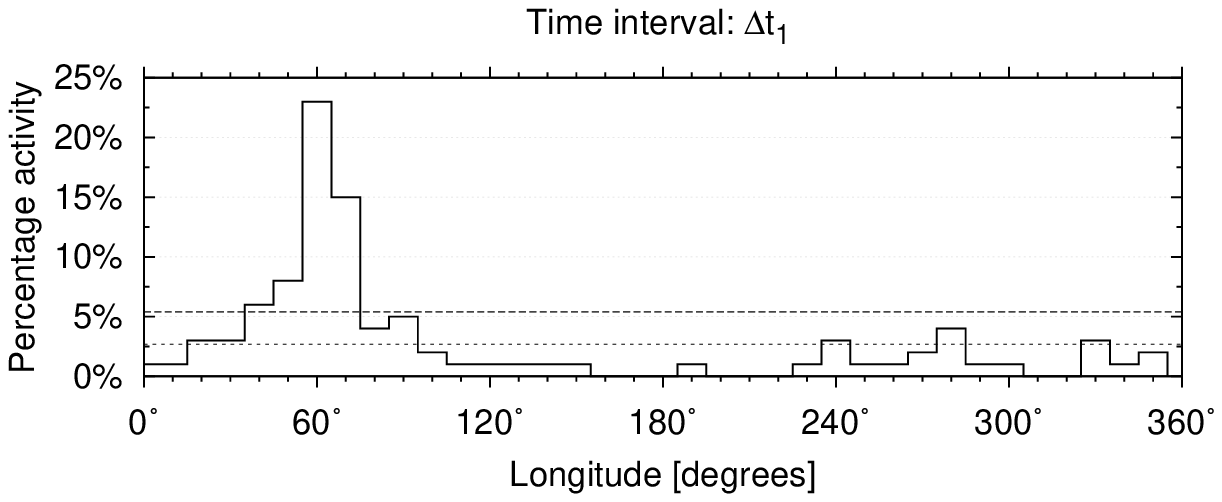}
               \hspace*{-0.03\textwidth}
               \includegraphics[width=0.515\textwidth,clip=]{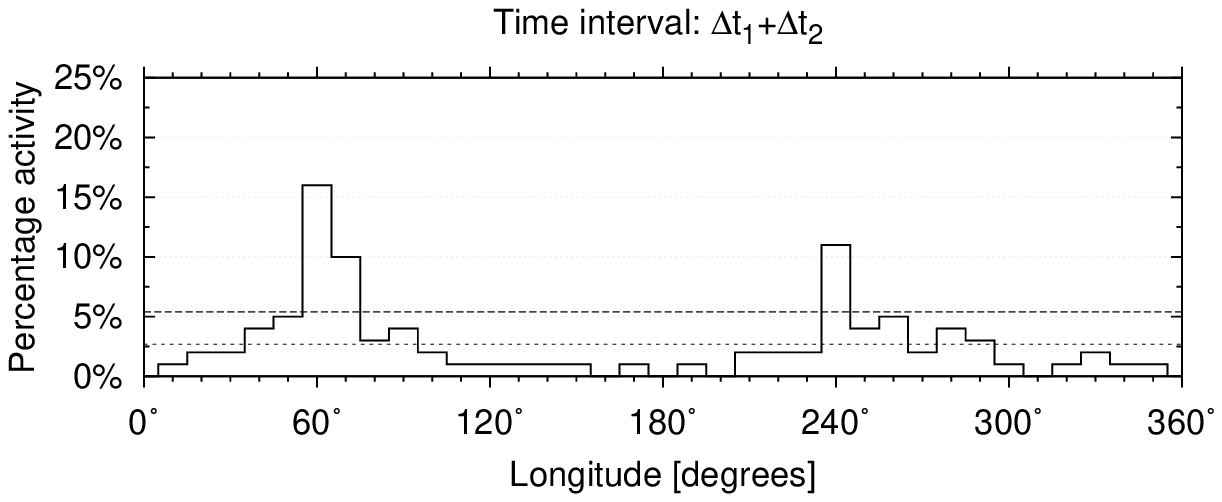}
                }
 \centerline{\hspace*{0.015\textwidth}
               \includegraphics[width=0.515\textwidth,clip=]{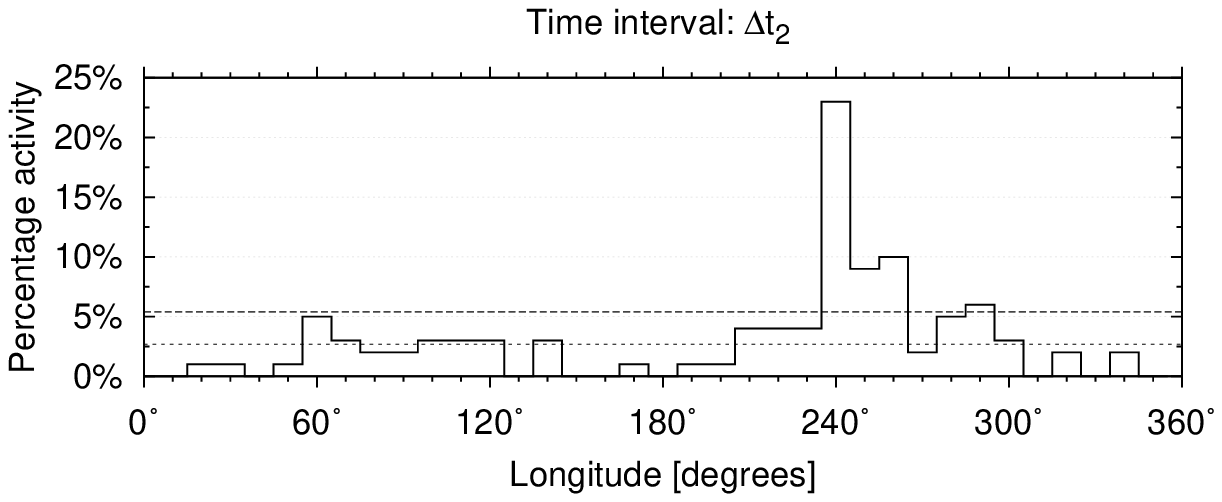}
               \hspace*{-0.03\textwidth}
               \includegraphics[width=0.515\textwidth,clip=]{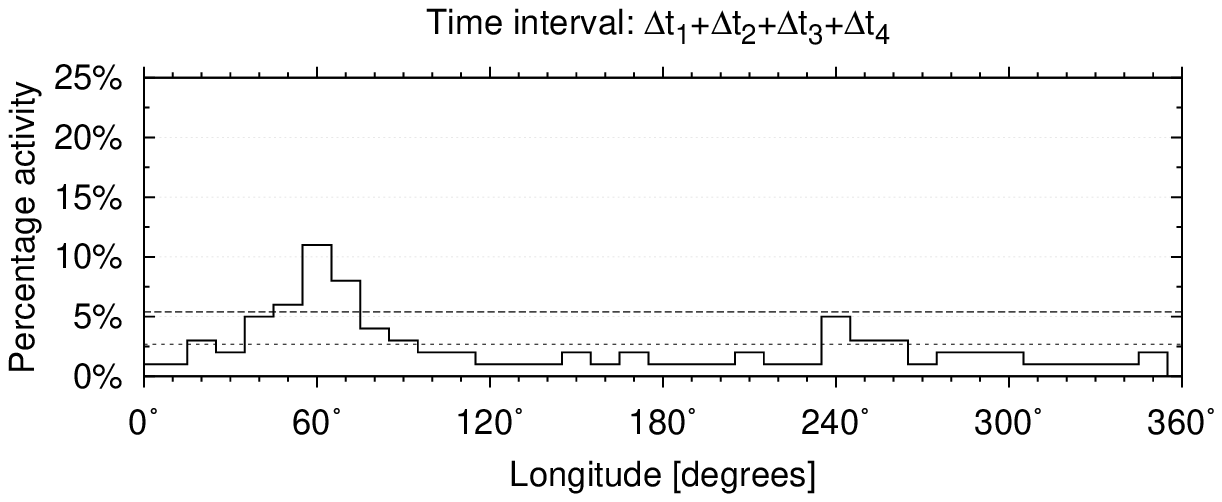}
                }
 \centerline{\hspace*{0.015\textwidth}
               \includegraphics[width=0.515\textwidth,clip=]{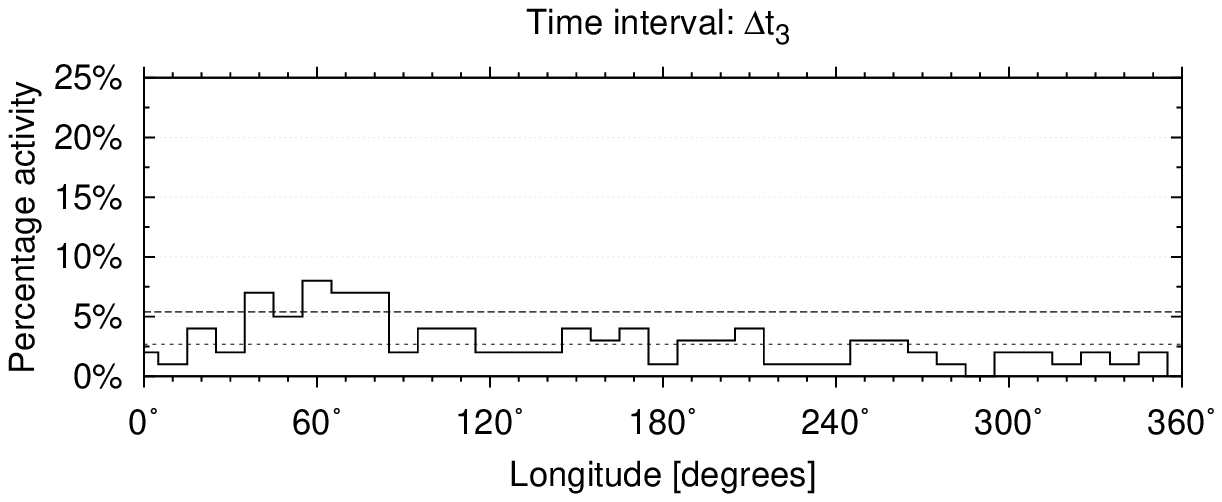} 
               \hspace*{0.494\textwidth}
              }
 \centerline{\hspace*{0.015\textwidth}
               \includegraphics[width=0.515\textwidth,clip=]{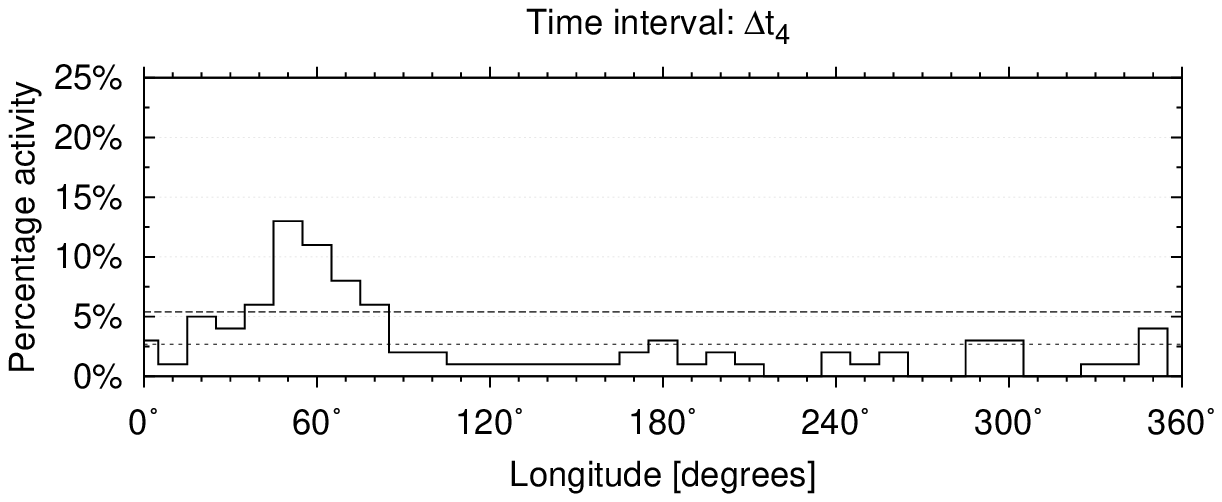}
               \hspace*{-0.03\textwidth}
               \includegraphics[width=0.515\textwidth,clip=]{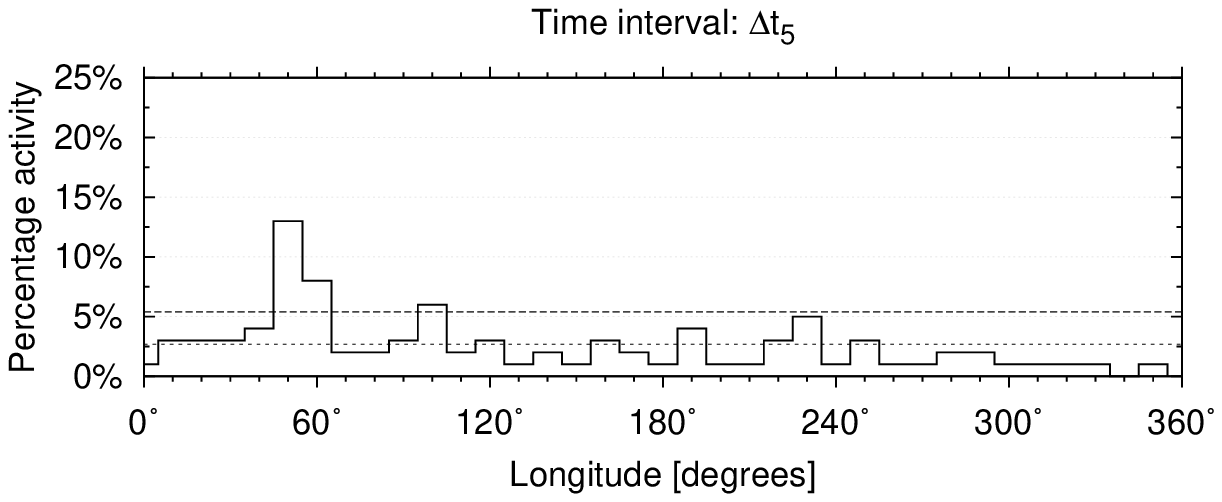}
                }

       \caption{Left panels: Longitudinal distribution of activity along the migration path in the northern hemisphere in the four time intervals indicated in Figure~\ref{north}. Right column; top panel: distribution in the united 1\,--\,2 intervals; second panel: distribution for the total length of the path; bottom panel: distribution in interval 5 indicated in the southern hemisphere, see Figure~\ref{south}. The horizontal dashed lines indicate the levels of $1\sigma$ and $2\sigma$ respectively.}

  \label{dist_NS}
  \end{figure}

Taking into account the migration of the active belts their longitudinal extension can be determined in a moving reference frame that moves along the parabola keeping its longitude at a constant value. In our moving reference frame the parabola is at $60^{\circ}$ longitude and the  $W_{i}$ values are calculated in $10^{\circ}$ bins for the time interval considered. In Figure~\ref{north} time intervals are selected in which the longitudinal distributions are different along the migration paths and therefore they are considered separately. The results are depicted in Figure~\ref{dist_NS}. It can be seen that the width of the active-longitude belt is about $60^{\circ}$ at $1\sigma$ level (3.5\%) and it is about $30^{\circ}$ at $2\sigma$ level (7\%) on average. 

The separation of the intervals 1\,--\,4 is not arbitrary. The first and second time intervals of Figure~\ref{dist_NS} show a flip--flop phenomenon in the northern hemisphere between Cycles 21 and 22. The upper right panel of the figure shows the distribution in the combined 1\,--\,2 interval. It contains two active longitudes at opposite positions with no activity halfway between them but the separate intervals 1 and 2 contain only one of them. Further consequences of the continuity through the interval 3 will be treated in Section~\ref{dynamics}.

\subsection{Another Identification of the Migration Path} 
  	    \label{path}

The migration path indicated in Figure~\ref{north} has been identified by visual inspection as a conspicuous formation in a noisy environment. The sharpness of the active zones presented in Figure~\ref{dist_NS} seem to corroborate the reality of the choice; nevertheless, different approaches may be necessary to justify this unexpected feature. 

The bottom lines of Figures~\ref{north} and ~\ref{south} show the diagrams of a parameter named SAL (Strength of Active Longitude) defined in the following way. The percentage of the total activity in longitudinal bins of $10^{\circ}$ has been plotted semiannually, the highest column has been chosen from each histogram, and the semiannual means have been subtracted from them. The differences have been plotted in semiannual resolution and the lowest lines of Figures~\ref{north} and ~\ref{south} show the domains of this histograms above 1$\sigma$. The values of SAL are shown between $12\%$ and $61\%$ in the northern hemisphere and between $9\%$ and $48\%$ in the southern hemisphere. The presence of the columns in consecutive semiannual intervals may be regarded as a necessary condition for the existence of any active longitude, the intervals of missing columns may be disregarded.

\begin{figure}

 \centerline{\hspace*{0.015\textwidth}
               \includegraphics[width=0.51\textwidth,clip=]{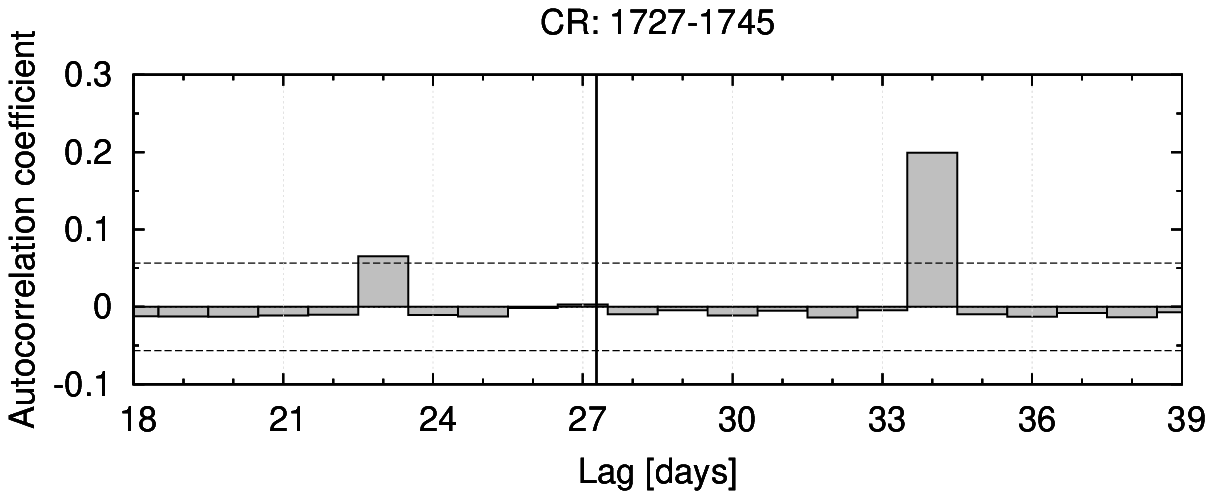} 
               \hspace*{0.49\textwidth}
              }

 \centerline{\hspace*{0.015\textwidth}
               \includegraphics[width=0.51\textwidth,clip=]{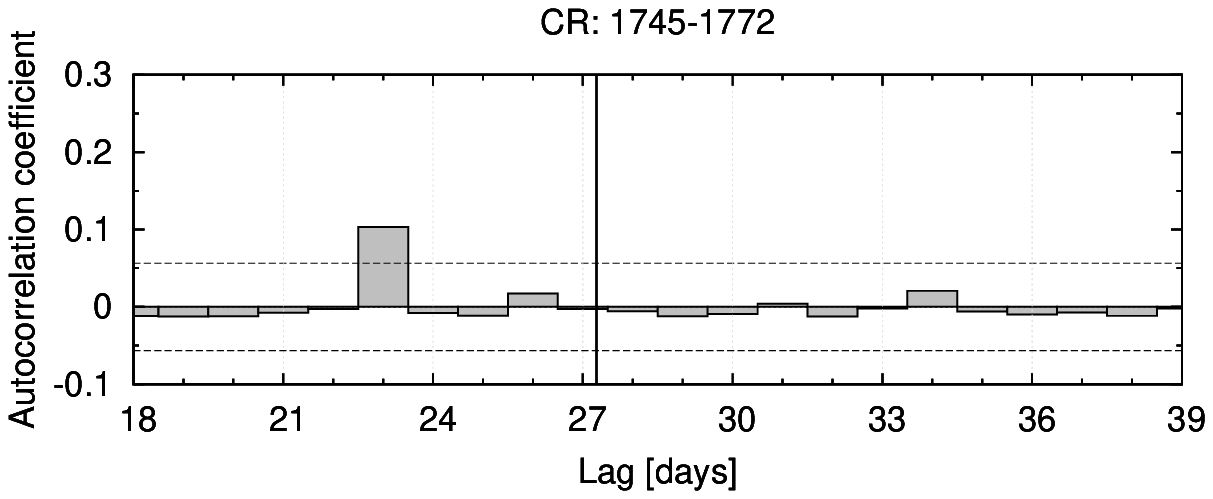}
               \hspace*{-0.03\textwidth}
               \includegraphics[width=0.51\textwidth,clip=]{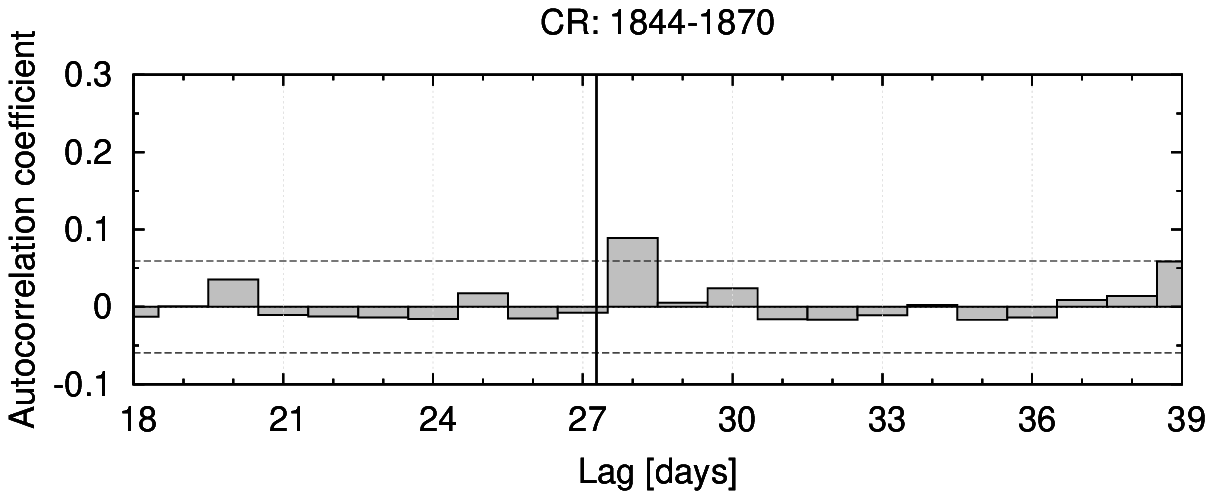}
                }

\centerline{\hspace*{0.015\textwidth}
               \includegraphics[width=0.51\textwidth,clip=]{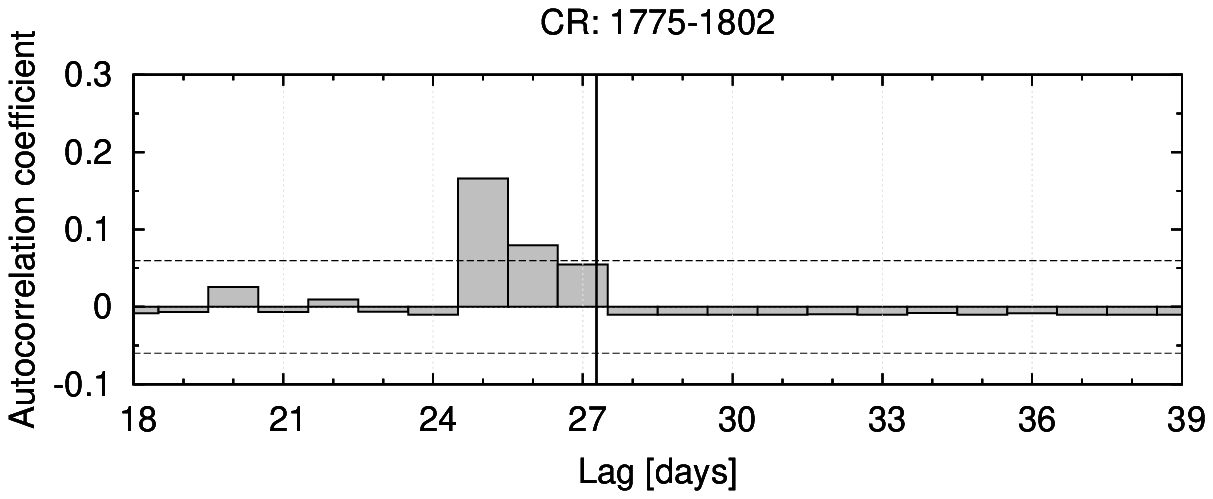}
               \hspace*{-0.03\textwidth}
               \includegraphics[width=0.51\textwidth,clip=]{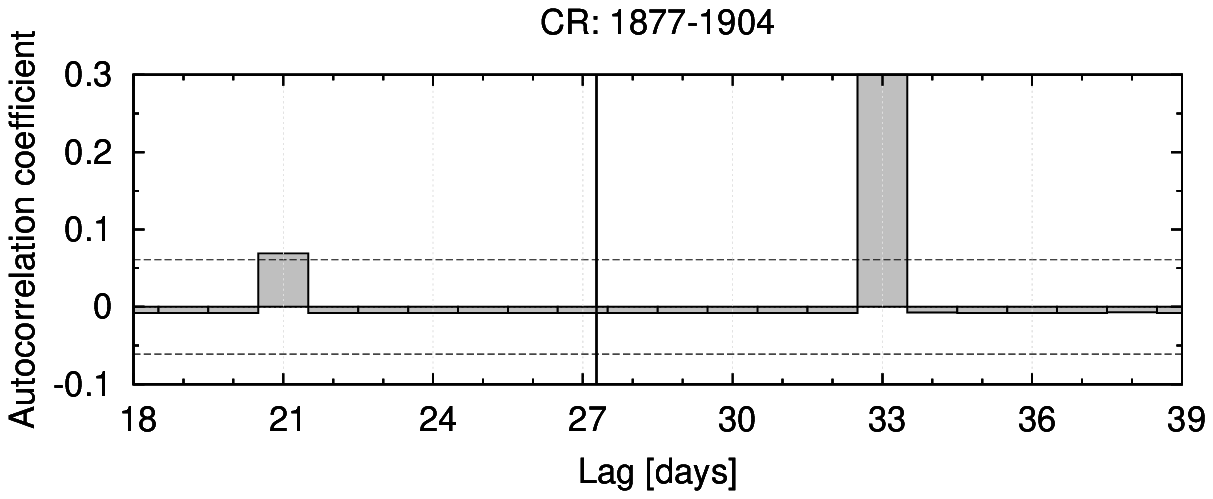}
              }

      \caption{Autocorrelations of five intervals to follow the variation of the rotation rate between the following Carrington Rotations: 1727\,--\,1745 (upper left, corresponds to interval a in Figure~\ref{north})  1745\,--\,1772, (middle left, interval 1),  1775\,--\,1802, (bottom left, the flip-flop event), 1844\,--\,1870, (upper right, interval 3, when the Carrington rate dominates), 1877\,--\,1904 (lower left, interval 4). The horizontal scale shows the synodic rotation rates, the vertical line marks the Carrington rate.}

 \label{rot}
 \end{figure}

The other method investigates the apparently varying rotation rate. As was summarized in the introduction, different investigations found different rotation rates depending on the methods and time intervals. To check the reality of the path marked in Figure~\ref{north} the rotation rates have been determined in two-year intervals shifted by steps of one Carrington Rotation along the time axis. In each two-year interval a time series has been created from the daily total area of sunspots observed within a distance of ${\pm} 10^{\circ}$ from the central meridian. The autocorrelations of these time series have been computed in each two-year interval. Figure~\ref{rot} shows the autocorrelations in five selected intervals. The first one is shorter than two years, corresponding to the backward trend prior to the parabolic path; the rest of them correspond to the intervals 1\,--\,4 indicated in Figure~\ref{north}. It can be seen how the rotation rate of the active-longitude zone varies with respect to the Carrington rate during the considered migration, or in other words, how the steepness of the path varies in the diagram of Figure~\ref{north}. The apparent variation of the rotation rate corresponds to the migration of the active longitude, and this correspondence can be regarded as the sufficient condition of the existence of a migrating active longitude. In the diagrams of the above defined SAL parameter the columns filled with gray color belong to those time intervals in which the autocorrelation values of the relevant rotation rates were significant, otherwise the column is white. This last case means that there are significant activity concentrations at certain longitudes temporally, however, these longitudes do not migrate smoothly, they can have jumps to distant locations. 

In the present work those time intervals are considered in which the above formulated necessary and sufficient conditions were mostly present simultaneously. This sampling principle allowed to select the time interval between Carrington Rotations 1708\,--\,1910 in the northern hemisphere, where these criteria were satisfied in $84.4\%$ of the time. In the southern hemisphere the two criteria were satisfied between Carrington Rotations 1990\,--\,2110 in $77.7\%$ of the time. The simultaneous fulfilment of these criteria was restricted to smaller fractions of other intervals.  This was the case in $23.5\%$ of the time between Carrington Rotations 1980\,--\,2110 in the northern hemisphere, and in $64.8\%$ of the time between Carrington Rotations 1700\,--\,1920 in the southern hemisphere, therefore they were not considered. These fractions are demonstrated in the bottom lines of Figures~\ref{north} and ~\ref{south}.

The presented shape of path may appear to be unexpected or even curious but in fact this article is not the first presentation of this sort of migration. Figure 5 of \inlinecite{Juckett06} presents practically the same shapes, the forward--backward migration within a couple of years and then an abrupt V-shaped turn to move forward again. The similarity is recognizable, {\it mutatis mutandis}, by taking into account the substantial differences between the two methods. 

It should be admitted that the parabolic fitting may be regarded as an assumption about the nature of the migration in spite of our original intention to avoid any pre-assumptions introducing subjective restrictions into the analysis. In the present case, however, the parabola has been chosen as the simplest function for this kind of shape; it does not imply any underlying physics at the moment. A fitting attempt with a section of a sinusoidal curve resulted in higher root-mean-squared error.

\subsection{Latitudinal Relations} 
  	    \label{latitudes}

\begin{figure}
  \centerline{\includegraphics[width=0.55\textwidth,clip=]{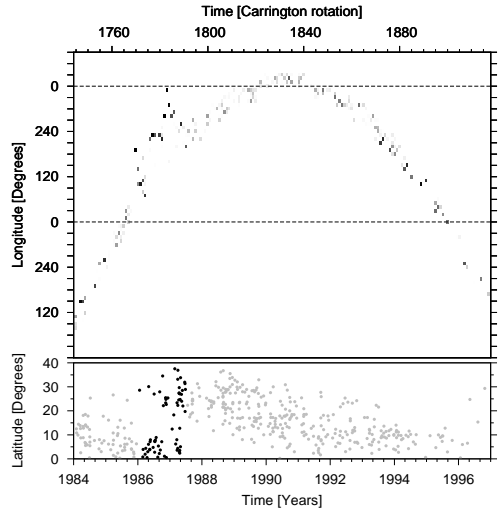}}

      \caption{The $60^{\circ}$-wide belts along the main and secondary paths in the time--longitude diagram (top panel). The Sp\"orer diagrams of the active regions within the paths (bottom panel). The main and secondary paths are distinguished by gray and black marks. }

  \label{jump}
  \end{figure}

The flip-flop phenomenon in the northern hemisphere happens between Cycles 21 and 22.  This is a jump of dominant activity from the main path to a secondary path at the opposite side, $180^{\circ}$ apart and back to the main path in the third interval. To examine the latitudinal relationship of these jumps, the Sp\"orer diagrams of Cycles 21 and 22 have been plotted in the northern hemisphere along with a $60^{\circ}$ wide belt along the path of enhanced activity, {\it i.e.} in the secondary path during interval 2, see Figure~\ref{jump}. Only the activity within these belts was considered. The activity centers of the main and secondary paths are distinguished by gray circles and black squares respectively. It can be seen that the flip-flop event happens at the change of the consecutive cycles and both cycles with their distant latitudes are involved in the activity of the secondary path.

\subsection{Dynamics of Emergence in the Active Belt} 
  	    \label{dynamics}

\begin{figure}
 \centerline{\hspace*{0.02\textwidth}
               \includegraphics[width=0.515\textwidth,clip=]{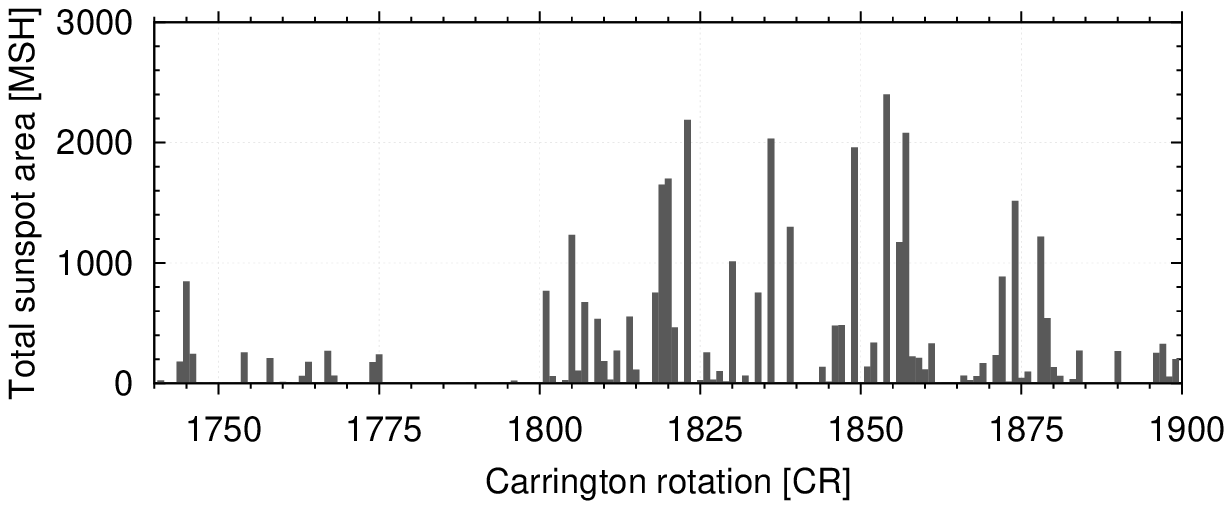}
               \hspace*{-0.03\textwidth}
               \includegraphics[width=0.515\textwidth,clip=]{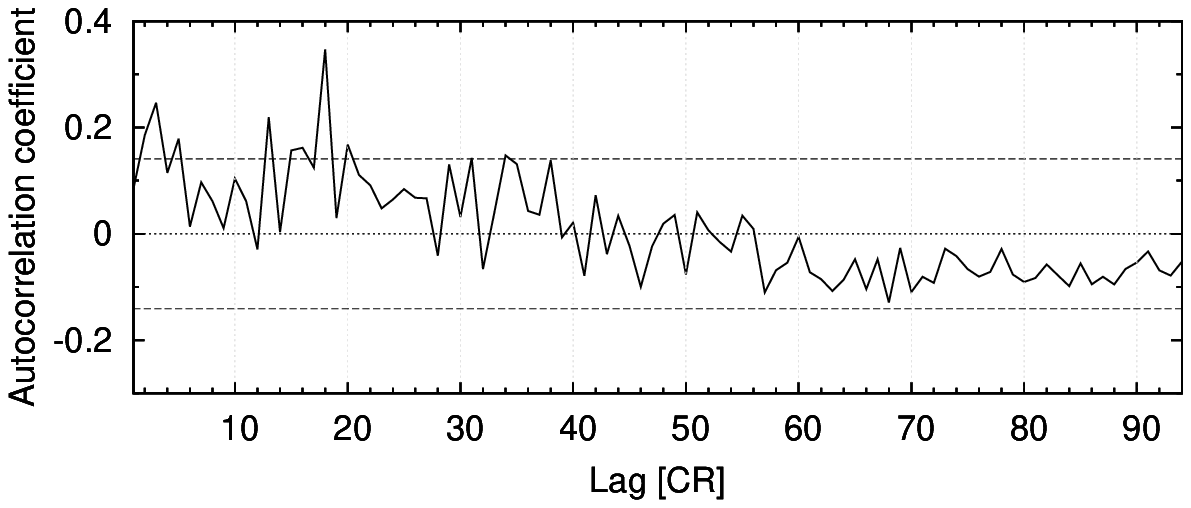}
              }

 \centerline{\hspace*{0.02\textwidth}
               \includegraphics[width=0.515\textwidth,clip=]{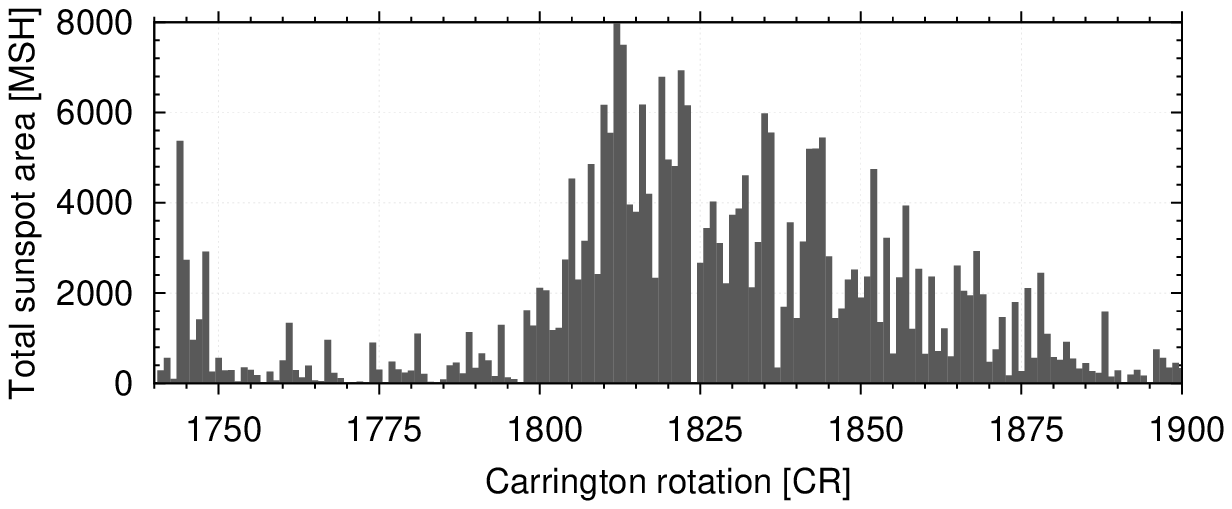}
               \hspace*{-0.03\textwidth}
               \includegraphics[width=0.515\textwidth,clip=]{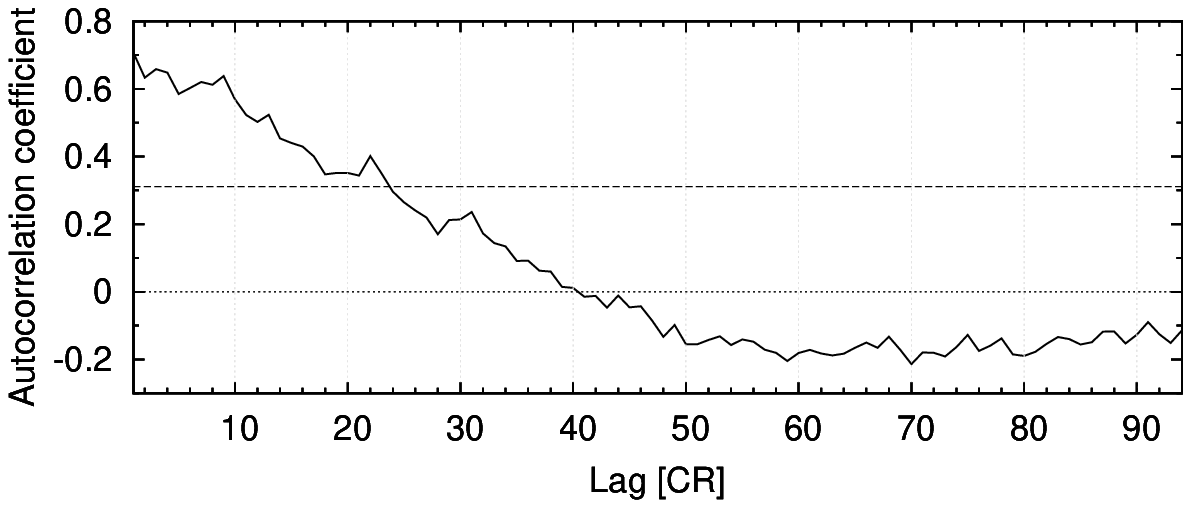}
               }

 \caption{ Upper row. Left panel: activity recorded by Carrington Rotations between rotations 1740\,--\,1900 within the $30^{\circ}$-wide longitudinal band around the parabola of Figure~\ref{north}; right panel: autocorrelogram of this data series with the confidence limits, the peak at 18 Carrington Rotation corresponds to 491 days or 1.345 years. Lower row. Left panel: total activity recorded by Carrington Rotations in the same time interval, right panel: autocorrelogram of this data series.}

 \label{correlogram}

\end{figure}

The variability of active-region emergence may also give some information about the nature of the belts of enhanced activity; this has been examined by autocorrelation analysis. We considered the $30^{\circ}$-wide band around the main path for Carrington Rotations 1740\,--\,1900. The upper panel of Figure~\ref{correlogram} shows the fraction of activity in these longitude intervals by rotation (top panel), the plotted values are the sums of the largest observed areas of all sunspot groups reaching their maxima in the given rotations. The lower panel of Figure~\ref{correlogram} shows the autocorrelation of this dataset; it has a single significant peak at the $18^{th}$ rotation which corresponds to about 1.3 years, more precisely 491 days or 1.345 years. 

To check whether this peak is an overall feature or it only belongs to the belt of enhanced activity, the same diagrams have been depicted for the sums of maximum sunspot areas on the entire Sun for the same rotations, see Figure~\ref{correlogram}. The $\approx$1.3-year peak disappeared from the autocorrelogram so this period can be attributed to the active longitude belt.

\section{Discussion}

It should be admitted that active longitudes cannot be identified reliably in a substantial fraction of the examined time period, so the following considerations are only related to those stripes of longitude which exhibit enhanced activity and continuous migration for at least two years according to the presented methods.

To interpret the migration of active longitudes a specific ``dynamic reference frame" has been put forward by \inlinecite{Usoskin05}. They assumed that the source region of the emerging fields rotates differently and by finding the appropriate constants in the rotation profile, the active longitudes can be localized. Their methodology was criticized by \inlinecite{Pelt05} and \inlinecite{Pelt06} but they maintained their concept and basic statements after correcting their formula \cite{Usoskin07}. They, among others, reported a clear cyclic behaviour. They evaluated the data by applying numerical criteria.

However, closer scrutiny of Figures~\ref{north} and ~\ref{south} shows that the migration paths of the active belts do not follow the shape of the 11-year cycle. The forward motion in Figure~\ref{north}  (interval 1) takes place during the declining phase of Cycle 21, as also observed on a restricted interval by \inlinecite{Bumba2000}, while the receding motion coincides with the declining phase of Cycle 22. However, no similar migration pattern can be recognized during Cycle 23. These figures show that differential rotation can hardly be involved in the migration of active longitudes, with neither solar nor with anti-solar profile. \inlinecite{Balthasar07} did not find signatures of differential rotation either. Juckett (\citeyear{Juckett06}, \citeyear{Juckett07}) applied surface spherical harmonics to the surface activity patterns; the variation of the phase information of this analysis describes the longitudinal migration of the active longitude. The results of these works also did not support the role of differential rotation.

Considering that the arc of the path extends to two cycles, one can only speculate about a possible impact of the Hale cycle because the turn-around point takes place at around the polarity change of the global dipole field. This issue needs more extended studies. 

The active-longitude belts also exhibit considerable temporal variation. The width of the belt may be as narrow as $20^{\circ}-30^{\circ}$ during moderate activity levels but around maximum it is dispersed as can be seen in Figure~\ref{dist_NS}. This narrow extension of the active longitude and the shape of its path in the longitude--time diagram can only be detected with the present resolutions: $10^{\circ}$ in longitude and one Carrington Rotation in time. It should be noted that the active longitudes are not always identifiable and the patterns of northern and southern hemispheres are not identical.

The 1.345-year period of spot emergence in the active belt may indicate its depth. \inlinecite{Howe2000} detected a radial torsional oscillation at the tachocline zone with this period during 1995\,--\,2000. The present study makes a distinction between the regions within and out of the active longitudinal region which is considered in a fairly narrow belt of $30^{\circ}$ along the parabola-shaped path in Figure~\ref{north}. The 1.3-year period is remarkable within the belt but it is absent, or overwhelmed, in the entire material. The presence of the $\approx$1.3-year period within the active belt allows the conjecture that the active longitudes may be connected to a source region close to the tachocline zone. This would be in accordance with \inlinecite{Bigazzi04} who argued that the non-axisymmetry can only remain at the bottom of the convective zone.

The flip-flop phenomenon was firstly observed by \inlinecite{Jetsu93} on the active star FK Comae Berenices, further analysed by \inlinecite{Olah06}. Figure~\ref{north} also shows two jumps between the main and secondary belts -- there and back. No regular alternation can be observed between the two regions in the considered time interval; similarly, \inlinecite{Balthasar07} did not find periodicity in the flip-flop events. What is even more important, the activity in the secondary belt takes place at the time of the exchange between Cycles 21 and 22, and furthermore, the active regions of this longitudinal belt are represented in both the ending and beginning cycles, {\it i.e.} at distant latitudes. This is a further hint that the migration of the active belt cannot be modeled with a differentially rotating frame. Figure~\ref{jump} gives the impression that the disturbance releasing the field emergence may be either a meridional feature, or a phenomenon affecting the field emergence in a broad latitudinal belt. In other words, the cause of the enhanced activity does not seem to belong to the toroidal field because it affects both of the old and new toruses simultaneously.

The literature provides quite different values for the rotation rate of the source region of active longitudes; some of them were cited in the introduction. In certain cases, some values can be identified as temporally detectable periods influenced by the actual forward or backward motion of the active longitude. Taking into account the steepness of the parabola-shaped path in Figure~\ref{north} during the decaying phases of Cycles 21 and 22 the active longitude migrates around the Sun forward and backward during about 34 rotations. This means a virtual synodic rotation period of 26.5 days in Cycle 21 and 28.1 days in Cycle 22. The reported periods between these values may be connected to the migration of the active longitude. \inlinecite{Jetsu97} also reported a 26.722-day period on the northern hemisphere from flare data but their 22.07-day period cannot be identified at the surface, as they also found. \inlinecite{Vernova04} reported a rigid rotation with the Carrington period; this cannot be confirmed with the recent findings.

The concept of a relic magnetic field investigated by \inlinecite{Olemskoy07} depends on the detectability of the 28.8-day period that corresponds to the rotation period of the radiative zone. This has not yet been detected in the surface data, and it is also missing from the present distributions. The relic magnetic dipole field is conjectured in the radiative zone and if its axis does not coincide with that of the convection zone magnetic field then their common effect could be a non-axisymmetric activity. This would imply a varying non-axisymmetry from cycle to cycle or the distinction between odd and even cycles because the polarity of the relic field is supposed to be constant. 

Another possibility could be a certain disturbance of the meridional flows that could distort the toroidal flux ropes resulting in their buoyancy and emergence. This disturbance could have a counterpart at the opposite side of the Sun, {\it i.e.} at the secondary belt of active longitude. The interaction of active regions and meridional motions has been investigated by \inlinecite{Svanda08} in an opposite causal order; the presence of the active region was considered as an obstacle for the flow. \inlinecite{Gonzalez08} also found modified meridional flows at the active regions that remain detectable even after the decay of the active regions. Longitudinal inhomogeneities of the meridional streams have not yet been reported. In the present data, the only intriguing coincidence is the forward migration in the northern hemisphere during negative magnetic polarity of the northern pole and backward migration after the change of polarity. 

\inlinecite{Dikpati05} published theoretical results on the active longitude problem. They concluded that the MHD shallow-water instability could produce bulges in the toroidal field that may be preferred sites to form magnetic loops rising to the surface. The $\approx$1.3-year periodicity of the activity within the active belt proper (Figure~\ref{correlogram}) could be a hint at this mechanism. To reproduce the migration pattern obtained, this mechanism would also have to imply a wave-like displacement of the bulge along the torus forward and backward depending on the polarity of main dipole field.

\section{Summary}

i) For the major fraction of dataset no systematic active longitudes were found. Sporadic migration of active longitudes was identified only for Cycles 21\,--\,22 in the northern hemisphere and Cycle 23 in the southern hemisphere. The following conclusions are related to the minor fraction of the data, when active longitudes can be identified at all, and not to the entire analyzed time period, they cannot be regarded as general features of sunspot emergence.

\noindent
ii) The active-longitude regions migrate in longitude forward and backward in the Carrington system independently from the 11-year cycle time profile. 

\noindent
iii) The half-width of the active longitudinal belt is $20^{\circ}-30^{\circ}$ during moderate activity but it is much less sharp at maximum activity. 

\noindent
iv) Flip-flop variations may occur but not with a regular period. 

\noindent
iv) Differential rotation is probably not involved in the migration of active longitudes. 

\noindent
vi) The active-region emergence exhibits a $\approx$1.3 year variation within a $30^{\circ}$-wide belt around the active longitude which is absent in the entire activity.

These results were obtained for a limited time interval of a few solar cycles and, bearing in mind uncertainties of the migration path definition, are only indicative. Further research based on an extended data set is needed because the present work is restricted to the DPD era: 1977\,--\,2011. The study of the the same features is under way on a longer dataset; the results will be published in a subsequent article.

\begin{acks}

The research leading to these results has received funding from the European Commission's Seventh Framework Programme (FP7/2007-2013) under the grant agreement eHEROES (project n° 284461, \href{http://www.eheroes.eu}{http://www.eheroes.eu}).

\end{acks}

\end{article} 

\end{document}